\mag=\magstephalf
\pageno=1
\input amstex
\documentstyle{amsppt}
\TagsOnRight
\interlinepenalty=1000
\NoRunningHeads
\pagewidth{14cm}
\pageheight{21.5cm}
\vcorrection{-1.0cm}
\hcorrection{-1.2cm}
\advance\vsize by -\voffset
\advance\hsize by -\voffset
\nologo


\font\twobf=cmbx12

\def\tvskip{\vskip 0.5 cm}

\define \RR{\Bbb R}

\define \TT{\Cal T}
\define \SS{\Cal S}

\define \CC{\Bbb C}
\define \bnabla{\overline{\nabla}}

\define \CCD{\Cal{D}}

\define \ii{\roman i}
\define \jj{\roman i}
\define \ee{\roman e}
\define \dd{\roman d}

\define \dalpha{{\dot \alpha}}
\define \dbeta{{\dot \beta}}

\def\mnarrower{\advance\leftskip by 50pt \advance\rightskip by 50pt}

{\centerline{\bf{Dirac Operator of a Conformal Surface 
Immersed in $\Bbb R^4$:  }}}
{\centerline{\bf{Further Generalized Weierstrass Relation}}}
\author
\endauthor
\affil
Shigeki MATSUTANI\\
2-4-11 Sairenji, Niihama, Ehime, 792 Japan \\
\endaffil
\endtopmatter


\subheading{Abstract}

In the previous report (J.~Phys.~A (1997) {\bf 30} 4019-4029), 
I showed that the Dirac operator defined over
a conformal surface  immersed in $\Bbb R^3$ is identified with the
Dirac operator which is generalized the Weierstrass-Enneper equation
and Lax operator of the modified Novikov-Veselov (MNV) equation. 
In this article, I determine the Dirac operator defined over
a conformal surface  immersed in $\Bbb R^4$, which is reduced
to the Lax operators of the nonlinear Schr\"odinger and
the MNV equations by taking appropriate limits.

\document

\tvskip
\centerline{\twobf \S 1. Introduction }
\tvskip

Investigation on an immersed geometry is current in
 various fields [1-6,8-25].
A certain class of the immersed surface in three dimensional 
space $\Bbb R^3$ is related to soliton theory.
The surfaces, {\it e.g.}, constant mean curvature surface,
constant Gauss curvature surface, Willmore surface 
and so on, can be expressed by soliton equation and are sometimes
 called soliton surfaces.
In the studies, Bobenko pointed out that an immersed surface in 
 $\Bbb R^3$
should be studied through the spin structure over it [1,2].
The auxiliary linear differential operator related to soliton 
surface should be 
regarded as an operator acting the spin bundle on the surface.

Recently Konopelchenko discovered  a Dirac-type operator defined 
over a conformal surface immersed in $\Bbb R^3$  which completely 
exhibits the symmetry of the 
immersed surface itself [3-4]. The Dirac-type differential equation
 for an conformal
surface immersed in $\Bbb R^3$ is given as
$$
    \partial f_1 = p f_2, \quad \bar \partial f_2= -p f_1,
        \tag 1-1
$$
where $p:= \frac{1}{2}\sqrt{\rho} H$,
$H$ is the mean curvature of the surface parameterized
 by complex $z$ and $\rho$ is  the conformal metric induced from 
 $\Bbb R^3$. 
(1-1) is a generalization of the Weierstrass-Enneper equation for 
the minimal surface [3-6].
This equation is sometimes called as generalized Weierstrass equation.
As the geometrical interpretation of the modified Novikov-Veselov 
(MNV) equation [8],  Konopelchenko  and Taimanov 
studied the dynamics of the surface obeying the MNV equation [3-6].

Independently  I proposed a  construction scheme of a
Dirac operator defined over an immersed object [8].
In a series of works [8-16], Burgess and Jensen and I
have been studying the Dirac operator in an  immersed object
 in $\RR^n$.
In refs.[8-13], I showed that the Dirac operator in an  
immersed curve in $\RR^n$
can be regarded as the Lax operator of soliton equation
while the extrinsic curvature of the immerse curve  obeys the soliton
 equation.
In other words, I obtained a natural spin structure over an immersed 
curve in $\Bbb R^n$.
In fact, I proved that the Dirac operator I obtained
is a canonical object and its analytic index agrees with the 
topological index 
of the immersed space [9,11].

After I proposed the construction scheme [8], Burgess and Jensen 
immediately
applied my construction scheme  to Dirac operator on a surface in 
$\RR^3$ [14].
In previous work [15], I show that if I restrict the surface to 
conformal one,
the Dirac operator turns out to be that of Konopelchenko (1-1) [2,3]
and by investigation of its quantum symmetry, its functional space 
 also exhibits the surface itself [16]. 
Hence it can be regarded that 
my construction scheme of the Dirac operator is canonical and 
gives structure over the immersed objects as Bobenko pointed 
out [1,2].

Recently using the Dirac operator in (1-1), Polyakov's extrinsic 
string [17] has been 
studied  by several authors [2-6,16,18-21]. As the Dirac operator 
in  (1-1) 
completely exhibits the symmetry of an immersed surface in $\Bbb R^3$,
in terms of the operator, one can investigate its symmetry.
In other words, the Dirac equation (1-1) can be interpreted as 
a generalization of
the Frenet-Serret relation of a space curve in the sense that 
the Dirac equation over an space curve can be regarded as "a square
 root" of the Frenet-Serret relation [8-13].

Investigation of a space curve itself is of concern as a polymer 
physics and as a geometrical interpretation of soliton theory [22-28].
Using the Ferret-Serret relation (or the Dirac equation over the
space curve) and isometry condition,
I proposed a calculation procedure to obtain an exact partition 
function of an closed space curve (elastica) in 
two dimensional space $\Bbb R^2$ based 
on a soliton theory [26].
There appears the modified Korteweg-de Vries (MKdV) equation as 
a virtual 
equation of motion or  thermal (quantized) fluctuation in the 
calculation of the partition function [26].
For the case of a space curve in  $\Bbb R^3$,
its partition function is  related to the complex MKdV equation 
and the nonlinear Schr\"odinger (NLS) equation [24,25,27].

Furthermore since there is natural relation between a surface and a 
space curve, the space curve interests us from the point of view of
 string theory [17].
Grinevich and Schmidt studied a closed space curve from  such 
viewpoint [28].
In fact the MNV equation can be regarded as a kind of
complexfication of the MKdV equation
and corresponding surface (Willmore surface) in $\Bbb R^3$
can be interpreted as a generalization of an elastica [3-7,21]; 
both codimension are one. 
 Using the correspondence and the Dirac operator in (1-1),
 I applied the calculation procedure of
the partition function of [26] to the Willmore surface which has 
Polyakov extrinsic action [21]. It means that I quantized the surface 
(a kind of string in the string theory)
 in $\Bbb R^3$ [21].
In the calculation, I used the Dirac operator in (1-1) to look into 
the symmetry
of the surface. Thus
the my calculation procedure of the partition function of an immersed
object is closely related to the construction scheme of the 
Dirac operator [15,16,21].

However a string in the string theory [17]
might be immersed in ten or twenty-four dimensional space.
Further it is expected that it will be quantized.
Thus I desire  the Dirac operator which expresses a surface 
immersed in higher dimensional space as a further generalization of 
the Dirac operator
of Konopelchenko or the generalized Weierstrass operator (1-1).  
In this article, I will 
give a Dirac operator defined on a conformal surface immersed in four 
dimensional space
$\Bbb R^4$ following the construction scheme I proposed in ref.[8].
 I conjecture that the Dirac operator which I will obtain might 
 exhibit the symmetry of 
the immersed object itself.
In fact, it becomes the Lax operator of NLS equation [11] and
that of the MNV equation [15] by taking appropriate limits.

Organization of this article is following. \S2 gives the review of 
the extrinsic
geometry of a conformal surface embedded in $\Bbb R^4$ and provides 
the notations
in this article. In \S3, I will confine a Dirac particle in the
 surface and determine the Dirac equation over the surface.
Using the complex representation, I will gives more explicit 
form of the 
Dirac operator in \S 4. \S 5 gives confirmation whether obtained 
Dirac operator is natural object by taking appropriate limits. 
I will discuss my result in \S 6.

\tvskip
\centerline{\twobf  \S 2  Geometry of a Curved Surface Embedded 
in $\RR^3$}
\tvskip


In this section, I will set up a geometrical situation of the system
which I will discuss,
and give the notations used in this article [11,16,21,28].
 I will deal with a conformal surface $\SS$ embedded in $\RR^4$
$$
	\Sigma \to \SS \subset \RR^4, \tag 2-1
$$
where $\Sigma = \CC/\Gamma$; $\CC$ is the complex plane and $\Gamma$ 
is a Fuchsian group.
Since  the imaginary time and the euclidean
quantum mechanics are very useful in the path integral method, 
I will deal with only the euclidean Dirac field in this article.
Furthermore, for sake of simplicity, I assume that
such a surface is embedded in $\RR^4$ rather than immersed for a 
while.

I assume that a position on a conformal compact surface
$\SS$  is represented using the affine vector 
${\bold x}(q^1,q^2)$$=(x^I)$
$ =(x^1,x^2,x^3,x^4)$ in $\RR^4$ and normal vectors of $\SS$ 
are denoted by ${\bold e}_3 $ and ${\bold e}_4 $.
Here $q^1$ and $q^2$ are natural parameters attached in the surface 
$\SS$.
The surface $\SS$ has  the conformal flat metric which is induced 
from the natural metric of $\Bbb R^4$,
$$
    g_{\alpha \beta} \dd q^\alpha \dd q^\beta=
      \rho \delta_{\alpha, \beta} 
    \dd q^\alpha \dd q^\beta . \tag 2-2
$$
I assume that the beginning parts of Greek indices 
$(q^\alpha,q^\beta,\dots)$
span from one to two  and adopt Einstein convention.
$\delta_{.,.}$ means the Kronecker delta matrix.

I will employ that the complex parameterization of the surface as,
$$
    z:=q^1 +\ii q^2 , \tag 2-3
$$
and
$$
    \partial := \frac{1}{2}(\partial_{q^1}-\ii \partial_{q^2}), \qquad
    \bar \partial:= \frac{1}{2}(\partial_{q^1}+\ii \partial_{q^2}) ,
    \quad \dd^2 q:=\dd q^1\dd q^2
    =: \frac{1}{2} \dd^2 z:=\frac{1}{2} \ii\dd z \dd \bar z.  \tag 2-4
$$
I sometimes express
it as $f=f(q)$ for a  real analytic function $f$ over $\SS$
but  I will use the notation $f=f(z)$ for a complex analytic function.

The moving frame over $\SS$ is denoted as
$$
     e^I_{\ \alpha}:= \partial_\alpha x^I,\quad
     e^I_{\ z}:=\partial x^I,  \tag 2-5
$$
where $\partial_\alpha 
:= \partial_{q^\alpha}:=\partial/\partial q^\alpha$.
Their inverse matrices are denoted as 
$e^\alpha_{\ I}$ and $e^z_{\ I}$.
The induced metric is expressed as
$$
  g_{\alpha\beta}=\delta_{I J}e^I_{\ \alpha}e^I_{\ \beta}, \quad
  \frac{1}{4}  \rho = <\bold e_z, \bold e_{\bar z}>
  := \delta_{I,J} e^I_{\ z} e^J_{\ \bar z}, \tag 2-6
$$
where $<,>$ denotes the canonical inner product in the euclidean 
space $\RR^4$. 

Since I will constraint a Dirac particle to be on  $\SS$ by taking 
an appropriate limit,
I can pay attention to only the vicinity of  $\SS$  
or a tubular neighborhood $\TT$ of the surface $\SS$  even 
before squeezing limit.
I will consider the properties of the tubular neighborhood $\TT$ 
and its affine structure in $\RR^4$.

I wish  that  geometry in the tubular neighborhood $\TT$ is expressed
using the variables of $\SS$. 
I will introduce  a general coordinate system $(q^1,q^2,q^3,q^4)$ 
besides  $(q^1,q^2)$
as a coordinate system of $\TT$. 
 The surface $\SS$ will be expressed as $q^3(\bold x)=q^4(\bold x)=0$.

Let the middle parts of the 
 Greek indices $(q^\mu,q^\nu,\cdots)$ used as curved system, 
 $\mu=1,2,3,4$.
Further I assume that the beginning parts of the Greek indices
  with dot $(q^\dalpha,q^\dbeta,\dots)$ run from three to four.

I will assume that the normal structure is trivial; the metric of 
the normal plane
is induced from $\Bbb R^4$ and vectors $\bold e_3$ and $\bold e_4$ are
orthonormal: $g_{\dalpha\dbeta}=\delta_{\dalpha\dbeta}$.
For later convenience, I also introduce the complex structure over the
normal plane,
$$
	\xi := q^3 + \jj q^4, \quad \bar \xi := q^3-\jj q^4. \tag 2-7
$$ 
 
The relation between the affine vector 
${\bold X}:=(X^1,X^2,X^3,X^4)$ in $\TT$
and natural coordinate systems  $q^\mu$ is given by
$$
    {\bold X}(q^\mu) = {\bold x}(q^\alpha) 
    + q^\dalpha {\bold e}_\dalpha . \ \  \tag 2-8
$$
This relation is uniquely determined  if 
$q^\dalpha$ is sufficiently small.

Next I will consider the extrinsic geometry of $\SS$ itself.
Using the moving frame $e^I_{\ \alpha}$,
I divide the ordinary derivative along $\SS$
into the horizontal and vertical
part; the horizontal part is written by $\nabla_\alpha$ defined as 
$$
 \nabla_\alpha {\bold b} := \partial_\alpha {\bold b}-
 <\!\partial_\alpha {\bold b},{\bold e}_\dalpha\!>{\bold e}_\dalpha ,
  \tag 2-9
$$ 
for a vector field ${\bold b}$. 
The two-dimensional Christoffel symbol 
$\gamma^\gamma_{\ \beta\alpha}$ 
attached on $\SS$ is thus defined as 
$$
	\nabla_\alpha{\bold e}_\beta
	=\gamma^\gamma_{\ \beta\alpha}{\bold e}_\gamma. \tag 2-10
$$
The second fundamental form is denoted as,
$$
          \gamma^\dalpha_{\ \beta\alpha}
        :=<\!{\bold e}_\dalpha,\partial_\alpha {\bold e}_\beta\!>,
        \quad
        \gamma^\xi_{\ \beta\alpha}
        =\frac{1}{2}(\gamma^3_{\ \beta\alpha}
        +\ii \gamma^4_{\ \beta\alpha})
                          . \tag 2-11
$$
On the other hand, 
the Weingarten map, 
$-\gamma^\alpha_{\ \beta \dalpha}{\bold e}_\alpha$, is defined by 
$$
	\gamma^\alpha_{\ \beta \dalpha}{\bold e}_\alpha
	:=\nabla_\beta{\bold e}_\dalpha , \qquad
	\gamma^\alpha_{\ \beta \dalpha}
	=<\!{\bold e}^\alpha,\partial_\beta{\bold e}_\dalpha\!> . 
	\tag 2-12
$$
Because of 
$\partial_\alpha\!\!<\!{\bold e}_\beta,{\bold e}_\dalpha\!>=0$,
 $\gamma^\alpha_{\ \beta \dalpha}$ is associated 
 with the second fundamental form through the relation,
$$
        \gamma^\dalpha_{\ \beta\alpha} = 
            -\gamma^\gamma_{\ \dalpha\alpha}g_{\gamma\beta}
         .  \tag 2-13
$$

Here I will choose the normal vectors $\bold e_\dalpha$ satisfied with,
$$
	\partial_\alpha {\bold e}_{\dalpha}
	=\gamma_{\ \dalpha \alpha}^{\beta} {\bold e}_{\beta}.
          \tag 2-14
$$
The derivative of a more general normal orthonormal base 
$\tilde{\bold e}_\dalpha$ is given as
$$
	\partial_\alpha \tilde{\bold e}_{\dalpha}
	= \tilde\gamma_{\ \dalpha \alpha}^{\beta} 
	\tilde{\bold e}_{\beta}.
         + \tilde\gamma_{\ \dalpha \alpha}^{\dbeta}
         \tilde{\bold e}_{\dbeta}  
          \tag 2-15
$$
instead of (2-14). 
From the orthonormal condition 
$\partial_\alpha < \tilde {\bold e}_{\dalpha},
\tilde{\bold e}_{\dbeta}>=0$,
the relations are obtained,
$$
    \tilde\gamma_{\ \dalpha \alpha}^{\dbeta}
    =- \tilde\gamma_{\ \dbeta \alpha}^{\dalpha}, \qquad
      \tilde\gamma_{\ \dalpha \alpha}^{\dalpha}\equiv 0 
      \ (\text{ not summed over }\dalpha). \tag 2-16
$$
In other words, there are only two parameters 
$\tilde\gamma_{\ 4\alpha}^{3}$ for $\alpha=1,2$.
Thus I will employ a SO(2) transformation so that 
I hold the relation (2-14);
$$
	\pmatrix \bold e_3 \\ \bold e_4 \endpmatrix
	  = \pmatrix \cos\theta &-\sin\theta \\
	   \sin\theta &\cos\theta \endpmatrix
	  \pmatrix \tilde{\bold e}_{3}\\ 
	  \tilde{\bold e}_{4} \endpmatrix,
	   \tag 2-17
$$
where
$$
	\theta := \int^{q^1} \dd q^1 
	\tilde\gamma_{\ 4 1}^{3}+\int^{q^2} \dd q^1 
	\tilde\gamma_{\ 4 2}^{3}.
	\tag 2-18
$$
This transformation is sometimes called as Hashimoto
 transformation [11,22,30].

As I prepared the languages to express the geometry of $\TT$, 
I will gives it.
In terms of $e^I_{\ \alpha}$,
 the moving frame of $\TT$ is described as,
$$
            E^I_{\ \mu} := \partial_\mu X^I ,     \tag 2-19
$$
$$
    E^I_{\ \alpha} = e^I_{\ \alpha} + 
            q^\dalpha \gamma^\beta_{\ \dalpha\alpha} e^I_{\ \beta}, 
            \quad  E^I_{\ \dalpha} = e^I_{\ \dalpha}.    
             \tag 2-20
$$
Its inverse matrix is denoted by $ (E^\mu_{\ I})$.
The induced metric of $\TT$ is given as
$$
    G_{\mu\nu} = \delta_{I J}E^I_{\ \mu} E^J_{\ \nu}  ,    
        \tag 2-21
$$
and is explicitly expressed as
$$ 
    \split
    G_{\alpha\beta}&=g_{\alpha\beta}+
        [\gamma_{\ \dalpha\alpha}^\gamma g_{\gamma\beta}+
         g_{\alpha\gamma}\gamma_{\ \dalpha\beta}^\gamma]q^\dalpha
         +[\gamma_{\ \dalpha\alpha}^\delta g_{\delta\gamma}
      \gamma_{\ \dbeta\beta}^\gamma]q^\dalpha q^\dbeta ,   \\
     G_{\dalpha\alpha}&=G_{\alpha\dalpha}=0, \\
    G_{\dalpha\dbeta}&= g_{\dalpha\dbeta}=\delta_{\dalpha\dbeta}.
                                                \endsplit \tag 2-22
$$
The determinant of the metric $G:={\roman {det}}(G_{\mu\nu})$ becomes
$$ 
    G=g \zeta\ ,\ \ \zeta:=
        (1+\roman{tr}_2(\gamma^\alpha_{\ 3\beta}) q^3
        +\roman{tr}_2(\gamma^\alpha_{\ 4\beta})q^4
                +K(q^3,q^4)), \tag 2-23
$$
where $g:={\roman {det}}_2(g_{\alpha\beta})$ and 
 $K(q^3,q^4)=\Cal O((q^3)^2, q^3q^4, (q^4)^2)$.
Here  $\roman{tr}_2$ is the two-dimensional trace over 
$\alpha$ and $\beta$. 
I will denote them as 
$$
H_1:=-\frac{1}{2}\roman{tr}_2(\gamma^\alpha_{\ 3\beta}), \quad
H_2:=-\frac{1}{2}\roman{tr}_2(\gamma^\alpha_{\ 4\beta}),
	 \tag 2-24
$$
and  introduce the "complex mean curvature",
$$
	H_\xi=H_1+ \ii H_2. \tag 2-25
$$
For an appropriate limit, it becomes ordinary mean curvature 
of one-codimensional [15] case
and for another limit, it becomes complex curvature of a 
space curve in $\Bbb R^3$
which is known as  curvature given by the Hashimoto 
transformation [11,22-24,27].

Using (2-22), the Christoffel symbols associated with 
the coordinate system 
of the tubular neighborhood $\TT$ is given as
$$
	\Gamma^\mu_{\ \nu \rho}:=\frac{1}{2}G^{\mu \tau}(
	G_{\nu \tau,\rho}+G_{\rho,\tau,\nu}-G_{\nu\rho,\tau}). 
	 \tag 2-26
$$
In terms of (2-26), the covariant derivative $\bnabla_\mu$ in 
$\TT$ is defined by,
$$
	\bnabla_\mu B_\nu:= \partial_\mu 
	B_\nu-\Gamma^\lambda_{\ \mu\nu}B_\lambda, \tag 2-27
$$
for a covariant vector $B_\mu$.

\tvskip
\centerline{\twobf \S 3 Dirac Field on a Surface Embedded in $\RR^4$}
\tvskip

As I set up the geometrical situation of my system, 
in this section I will consider the Dirac field 
 $\pmb{ \Psi}=:(\pmb{ \Psi}^1,\pmb{ \Psi}^2)^{T}$
defined in the tubular neighborhood   $\TT$ and confine it into $\SS$ by takin
g a limit [8-16]. 
 
I will start with the original lagrangian  given by, 
$$
   \Cal L \dd^4 x =  \bar{\pmb{ \Psi}}(x) \ii 
   (\Gamma^I \partial_I -m_{\roman{conf}}(q^{\dalpha}))
   \pmb{ \Psi}(x)\ \dd^4 x  , \tag 3-1
$$
where $\Gamma^I$ is the gamma matrix in the  $\RR^4$,
$\partial_I:=\partial/\partial x^I$ and 
$\bar {\pmb{\Psi}}={\pmb{\Psi}}^\dagger \Gamma^1$.
I assume that  $m_{\roman{conf}}$ has the form, 
$m_{\roman{conf}}(q^\dalpha):
=\sqrt{\mu_0^2+\omega^2((q^3)^2+(q^4)^2)} $ 
for a large $ \omega $ and $\mu_0$ with 
$\mu_0 \gg \sqrt{\omega}$ [13,15].

I will denote a  thin tubular neighborhood by $\TT_0$.  
I can approximately 
regarded $\TT_0$ as a trivial disk bundle; 
$\TT_0 \approx {\frak D}_{1/\omega}\times\SS$, 
${\frak D}_{1/\omega}:=\{ (y_1,y_2)\Bigr|\  |(y_1,y_2)|<1/\omega\}$.
As well as in refs.[8-16,31-33], by paying my 
attention only on the ground
state, the Dirac particle is approximately confined in the thin 
tubular neighborhood $\TT_0$.
Even though there is a U$(1)(\approx{\frak D}_{1/\omega}-\{0\})$
 symmetry around $\SS$, 
I will choose a trivial rotation
or constant function as a section of a sphere bundle 
U$(1)\times \SS$ because 
non-trivial rotational component of the energy is 
expected as order of $\omega$.
After taking squeezed limit, I can realize the 
quasi-two-dimensional subspace in $\RR^4$
and, by integrating the Dirac field along the normal direction,
express the system using the two-dimensional parameters $(q^1,q^2)$.
Then I will interpret the Dirac field as that over the surface 
$\SS$ itself [8-16].

Thus I will express the lagrangian in terms of the curved coordinate 
system  $q^\mu$.
For the coordinate transformation (2-8), the Dirac operator 
becomes [8,13,15]
$$
	\ii \Gamma^I \partial_I =\ii 
	\Gamma^\mu \partial_\mu, \tag 3-2
$$
and the spinor representation of the coordinate transformation 
is given as
$$
	\pmb{\Psi}(q)= \ee^{- \Sigma^{I J} 
	\Omega_{I J}} \pmb{\Psi}(x), \tag 3-3
$$
where $\Sigma^{I J}$ is the spin matrix,   
$$
	\Sigma^{I J}:=\frac{1}{2}[\Gamma^I,\Gamma^J], \tag 3-4
$$
and $\Sigma^{I J} \Omega_{I J}$ is a solution of the differential
 equation,
$$
 \partial_\mu \left(\Sigma^{I J} \Omega_{I J}\right)=\Omega_\mu, 
\qquad
  \Omega_\mu := \frac{1}{2} 
          \Sigma^{I J} E_I^{\ \nu}(\bnabla_\mu E_{J\nu}). 
          \tag 3-5
$$

Hence the lagrangian density (3-1) becomes
$$
 \Cal L \ \dd^4 x = 
  \bar{\pmb{\Psi}}(q)\ii (\Gamma^\mu D_\mu
   -m_{\roman{conf}}(q^\dalpha))\pmb{\Psi}(q)\ \sqrt{G}\ \dd^4 q, 
                   \tag 3-6
$$
where  $\Gamma^\mu:=\Gamma^I E_I^{\ \mu}$ and 
$D_\mu$ denotes the spin connection,
$$
    D_\mu:=(\partial_\mu + \Omega_\mu).  \tag 3-7
$$
Here I will note the relation for the normal direction,
$$
	D_\dalpha = \partial_\dalpha \quad \text{module} \ q^\dbeta .
	 \tag 3-8
$$

Since the measure on the curved system is given as,
$$
	\dd^4 x=\sqrt{G}\cdot\dd^4q, \tag 3-9
$$
and  $-\ii D_\dalpha$ is not hermite nor a momentum operator,
I redefine the field  as [8-16,30-33],
$$
		\Psi=\zeta^{1/2}\pmb{ \Psi} . \tag 3-10
$$
Then the lagrangian density (3-6) changes as,
$$
 \Cal L \ \dd^4 x =  
 \bar{\Psi}(q)\ii (\Gamma^\mu \Bbb D_\mu 
 -m_{\roman{conf}}(q^\dalpha))\Psi(q)\ \sqrt{g}\ \dd^4 q
    \ , \tag 3-11
$$
where
$$
	\Bbb D_\alpha:=D_\alpha-\frac{1}{4} 
	\partial_\alpha \log \zeta, \qquad
	\Bbb D_\dalpha := D_\dalpha + 
	\frac{H_{\dalpha-2} -\partial_\dalpha K}
	{1-2 H_1 q^3-2 H_2 q^4+K(q^3,q^4)}. \tag 3-12
$$
Due to (3-10), in the deformed Hilbert space spanned by $\Psi$, 
$-\ii D_\dalpha$ is a generator of the translation along the normal 
direction and thus it is the momentum operator.
Thus after squeezing limit, the normal direction is independent space
and can be regarded as a inner space.

More mathematical speaking, the quantum physics is based on tangent
 bundle of a manifold rather than the manifold itself. 
The original differential operator 
should be interpreted as an operator defined over
$T\Bbb R^4\approx \Bbb R^8$ rather than $\Bbb R^4$ [33].
More precisely speaking, in the case of the Dirac particle,
the Dirac operator acts upon the spin bundle over $\Bbb R^4$,
Spin($\Bbb R^4$) and its fiber has the structure of Clifford algebra 
induced from the tangent bundle $T\Bbb R^4$ [34].
The differential operator  $-\ii \partial_I$ is a generator of the 
translation $\Bbb R^4$ and should be regarded as a base of the 
fiber space of $T\Bbb R^4$ or Spin($\Bbb R^4$).
Hence after coordinate transformation from $x^I$ to $q^\mu$ 
by restricting
the manifold $\Bbb R^4$ to $\TT_0$, I interpret the Dirac operator
as an operator over the spin bundle Spin($\TT_0$) induced from
the tangent bundle $T\TT_0$.
The action of (3-11) decomposes the function space over 
the tangent space of $T\TT_0$ to that over 
the tangential and that over the normal direction for $\SS$. 
By (3-8) and (3-10), the normal part is asymptotically equivalent 
with that of $T \Bbb R^2$
in the sense of real analytic function. In other words,
$D_\dalpha$ becomes the base of the fiber space and due to the 
confinement potential, the function space along the normal direction
is effectively restricted as a set of compact support
functions over the fiber space 
${\frak D}_{1/\omega} \approx \Bbb R^2$ of  $\TT_0$. 
In the limit, $D_\dalpha$ behaves like $\partial_\dalpha$ which is a
translation operator
of flat space  ${\frak D}_{1/\omega} \approx \Bbb R^2$ of the 
fiber space of  $\TT_0$.
In other words, after squeezing limit, $T \TT_0$ is reduced to
$T\SS \times T {\frak D}_{1/\omega}$ as a trivial bundle over $T\SS$.
Then I can regard the normal direction $T{\frak D}_{1/\omega}$ 
 as an inner space over $T\SS$ [8-16,31-33].

Such inner space comes from
the symmetry of the euclidean space $\Bbb R^4$ and is
restricted by an immersed object.
Hence at least, in the cases of a space curve in $\Bbb R^n$ $n>1$,
and a surface in $\Bbb R^3$, the Dirac operator completely
exhibits the embedded symmetry [8-16].

After squeezing limit,
due to the confinement potential $m_{\roman{conf}}$, 
the Dirac field for the normal direction is asymptotically factorized
 and can 
be expressed by the modes classified by a non-negative integer $n$ 
[8-15].
Then there exists a mode  with the lowest energy
for normal direction and trivial rotation  such that
$$
	\Psi(q^1,q^2,q^3,q^4)\sim \sqrt{\delta(|q^\dalpha|)} 
	\pmb{\psi}(q^1,q^2) \tag 3-13
$$
and
$$
     (\Gamma^\dalpha\partial_\dalpha -m_{\roman{conf}}(q^\dalpha)) 
     \sqrt{\delta(q^\dalpha)} 
     \pmb{ \psi}(q^1,q^2)=m_0\sqrt{\delta(|q^\dalpha |)} 
     \pmb{\psi}(q^1,q^2) .
		\tag 3-14
$$

By restricting the function space of the Dirac field to the
 ground state for the normal direction ($n=0$), I will define
the lagrangian density on a surface $\SS$  [15],
$$
	\Cal L_{S}^{(0)} \sqrt{g} \dd^2 q 
	:=\left.\left(\int \dd q^3\dd q^4 
	\Cal L\ \sqrt{G}\right) \right|_{n=0} 
	\dd^2 q , \tag 3-15
$$
and then it has the form,
$$
    \Cal L_{S}^{(0)} \sqrt{g} \dd^2 q 
    =\ii \bar{\pmb{ \psi}}(\gamma^1 \CCD_1
    +\gamma^2\CCD_2+ H_1 \gamma^3 +H_2 \gamma^4 + m_0 )
    \pmb{ \psi}\ \sqrt{g} \dd^2 q  , \tag 3-16
$$
where I rewrite the quantities as 
$$e^I_{\ \alpha} \equiv E^I_{\ \alpha}|_{q^\dalpha=0}, \qquad
	\gamma^\mu(q^1,q^2):=\Gamma^\mu(q^1,q^2,q^\dalpha\equiv0), 
$$
$$
	\omega_\alpha(q^1,q^2)
	:=\Omega_\alpha(q^1,q^2,q^\dalpha\equiv0),\qquad 
	\CCD_\alpha:=\partial_\alpha +\omega_\alpha  .\tag 3-17
$$

Since a confined space is expressed by the two-dimensional parameter 
$(q^1,q^2)$ and can be regarded as a two-dimensional space, 
I can identify the confined space with the surface $\SS$
itself and then the differential
operator in (3-16) is interpreted as a Dirac operator in a surface
 $\SS$ embedded in $\RR^4$.
Hence the spin connection can be written as
$$
    \omega_\alpha := \frac{1}{2} \Sigma^{i j} 
    e_i^{\ \beta}(\nabla_\alpha e_{j\beta}),
                                        \tag 3-18
$$
where the indices of $i,j$ run from 1 to 2. 

It should be noted that the Dirac operator in (3-16) 
is not hermite in general as that in a space curve immersed in 
$\RR^4$ is neither [8-16].
It is natural because the extra term in the corresponding 
Schr\"odinger equation
in the surface $\SS$ behaves the negative potential [30-33]; 
roughly speaking 
the square root of the negative potential appears as a pure imaginary 
extra field  in the Dirac equation [8-16].

As I argued in ref.[15], an immersion can be approximately 
regarded as embedding 
an object in more higher dimensional space;
A loop soliton in $\Bbb R^2$ is mathematically realized as an 
immersed curve in $\Bbb R^2$ but is physically realized as a 
thin rod on a plane in $\Bbb R^3$,
whose overlap is negligible and their dimensions can be approximately
regarded as one and two of an immersed curve
in $\Bbb R^2$.
Thus I can generalize this result of an embedding surface in 
$\Bbb R^4$ to that of immersed case.

\tvskip
\centerline{\twobf \S 4 Dirac Operator on a Complex Surface
 Immersed in $\RR^4$}
\tvskip

Generally $m_0$ does not vanish, but I will neglect the mass $m_0$ 
hereafter as in refs.[8-16].
I am interested in the properties of the Dirac operator itself and
I will investigate the high energy behavior of the field with 
$m_0$, {\it i.e.}, the behavior of the  high energy for surface 
direction and the lowest energy of the normal direction
using the independency of the normal modes.

Since the  surface $\SS$ is  conformal, 
I will explicitly express (3-16).
Then the Christoffel symbol is calculated as [17],
$$
	\gamma^\alpha_{\ \beta \gamma}=
	\frac{1}{2} \rho^{-1}
	(\delta^\alpha_{\ \beta}\partial_\gamma \rho
	         +\delta^\alpha_{\ \gamma}\partial_\beta \rho
	           -\delta_{\beta \gamma}\partial_\alpha \rho) .
	      \tag 4-1
$$
I will denote a natural euclidean inner space by
$y^a,y^b,\cdots$, $(a,b=1,2)$.
Then the moving frame is written as,
$$
	e^a_{\ \alpha}=\rho^{1/2} \delta^a_{\ \alpha}, \tag 4-2
$$
and the gamma matrix is connected to the Pauli matrices $\sigma^a$,
$$
	\gamma^\alpha=e_a^{\ \alpha} \sigma^1 \otimes \sigma^a, 
	 \tag 4-3
$$
Thus the spin connection becomes
$$
	\omega_{\alpha}=- \frac{1}{4} \rho^{-1}\sigma^{ab}
	(\partial_a  \rho \delta_{\alpha b}
	-\partial_b  \rho \delta_{\alpha a}) , \tag 4-4
$$
where $\sigma^{ab}:=1\otimes [\sigma^a,\sigma^b]/2$.
The Dirac operator in (3-16) can be expressed as
$$
	\gamma^\alpha \CCD_\alpha
	=\sigma^1\otimes\sigma^a \delta^\alpha_{\ a}
		[\rho^{-1/2}\partial_\alpha + \frac{1}{2}\rho^{-3/2} 
		(\partial_\alpha \rho)]  .
		\tag 4-5
$$

Similar to (3-10), I will redefine the Dirac field in the surface 
$\SS$ as
$$
	\psi:=\rho^{1/2} \pmb{ \psi} \tag 4-6
$$
and then the Dirac operator (4-5) becomes simpler [15,17],
$$
	\gamma^\alpha \CCD_\alpha\pmb{ \psi}
	= \rho^{-1}\sigma^a \delta^\alpha_{\ a}\partial_\alpha\psi .
	\tag 4-7
$$

On the other hand, due to the properties of the Dirac matrix
and noting the fact that the normal direction should be regarded as
an inner space, $\gamma^{\dalpha}$ is represented as,
$$
         \gamma^3 =\sigma^1 \otimes \sigma^3,\quad  \gamma^4 
         =\sigma^2 \otimes 1.	  \tag 4-8
$$
 Since $\bar {\pmb{\psi}}\gamma^1\pmb{\psi}\sqrt{g} \dd^2 q$
is the charge density, I expect the relation,
$$
	\bar \psi= \psi^\dagger \gamma_1 
	=\psi^\dagger \sigma^1\otimes\sigma^1  \rho^{1/2} . \tag 4-9
$$

Accordingly the lagrangian density (3-16) is reduced to,
$$
    \Cal L_{S}^{(0)} \sqrt{g} \dd^2 q =
    \ii \bar{\psi}\rho^{-1/2}
    (\sigma_1\otimes \sigma^a\delta^\alpha_{\ a}\partial_\alpha
    + \rho^{1/2} H_1 \sigma_1\otimes \sigma_3 
    + \rho^{1/2} H_2 \sigma_2\otimes 1) \psi\  \dd^2 q 
     \tag 4-10
$$
Here I will denote the field as
$$
	\psi = \pmatrix \psi_{1+}\\ \psi_{1-} \\ \psi_{2+}\\ 
	\psi_{2-}\endpmatrix .
	\tag 4-11
$$
Using the complex parameterization of the surface (2-3) and (2-4),
 the lagrangian density (4-10) can be explicitly expressed.
 Then the equation of motion of the Dirac field in the conformal 
 surface $\SS$ immersed in $\RR^4$ is derived as
$$
        \pmatrix & &  p_c & \partial \\
                 &  & \bar \partial & -\bar p_c \\
                  \bar p_c & \partial & & \\
                  \bar \partial &  -p_c & & \endpmatrix
                   \pmatrix \psi_{1+}\\ \psi_{1-} \\ \psi_{2+}\\ 
                   \psi_{2-}\endpmatrix=0,
                   \tag 4-12
$$
where the "external" field $p_c$ is defined as,
$$
	p_c:=\frac{1}{2}  \rho^{1/2} H_\xi. \tag 4-13
$$
(4-12) is the Dirac equation canonically defined over the 
conformal surface $\SS$ immersed in $\Bbb R^4$.
This equation is a generalization of that of a space curve immersed 
in $\Bbb R^3$ [8-13] and that of a conformal surface immersed in 
$\Bbb R^3$ [14-16] following my procedure [8-16].

\tvskip
\centerline{\twobf \S 5  Some Limits} 
\tvskip

In this section, I will check whether the obtained Dirac operator 
in (4-12)
naturally behaves after taking some limits.

\subheading{\S 5-1 Reduction to a space curve}

First I will consider a space curve $\Cal C$ in $\Bbb R^3$, 
which is parameterized by
$q^1 \in S^1$. Its codimension is two.
Due to the codimensionality, this situation resembles a 
surface immersed in $\Bbb R^4$,
which I dealt with until the previous section.
 I can extend $\Cal C \subset\Bbb R^3$
to $\SS \subset \Bbb R^4$ by multiplying a trivial 
line manifold $\Bbb R$;
 $\Cal C\times \Bbb R$ can be regarded as a surface $\SS$.
This surface is not compact and is contradict in my assumption 
in (2-1).
However it is not difficult to see that
 the assumption can be  extended to this case.
 
Hence I can regard this situation as a special case of the 
situation which
I argued until the previous section. 
I explicitly represent the "mean curvature" in (2-24) 
and metric in (2-6),
$$
	H_1= -\frac{1}{2} \text{tr}_2(\gamma^1_{\ 31}),\quad
	H_2= -\frac{1}{2} \text{tr}_2(\gamma^1_{\ 41}), \quad
	\rho=1. \tag 5-1
$$
They are a half of curvatures under the Hashimoto representation 
of the space curve
and $ H_\xi$ is a complex curvature. 
Vortex soliton obeys the NLS equation and whose dynamics is 
expressed in terms
of $H_\xi$ [22-24],
$$
	\ii \partial_t H_\xi +\frac{1}{2} \partial_1^2 H_\xi 
	+|H_\xi|^2 H_\xi =0 .
	\tag 5-2
$$

Now I will consider (4-12) in this case.
 Then there is a trivial solution $\psi=\psi(q^1)$
and $\partial_2 \psi=0$.
Noting $\partial f(q^1)=\partial_1 f(q^1)/2$, (4-12) reduces to
$$
	L= \pmatrix   H_\xi& \partial_1\\  
	\partial_1 & -\bar H_\xi  \endpmatrix.
	\tag 5-3
$$
which is one of its Lax operator of the NLS equation [34].
I obtained (5-3) by considering the Dirac operator over an immersed 
curve in $\Bbb R^3$ in refs.[10,11]. 
Further (5-3) can be regarded as the Frenet-Serret equation by
inverse of the Hashimoto transformation [10,11].
It is a remarkable fact that the Dirac operator in (4-12) reproduces
the Dirac operator in refs.[10,11] and is associated with the 
vortex soliton.

\subheading{\S 5-2 Reduction to a surface in $\Bbb R^3$}

Next I will constrain the surface $\SS$ to be immersed in $\Bbb R^3$,
which Konopelchenko and Taimanov dealt in refs.[3-6] and I considered
in the previous paper [15].

I will suppose that the direction ${\bold e}_4$ is identified with the
direction of $x^4$ and ${\bold e}_4$ geometry becomes trivial;
$$
	H_2= 0 \tag 5-4
$$
Then the external field is obtained as,
$$
	p_c \equiv p_1:= \frac{1}{2}\sqrt{\rho} H_1  \tag 5-5
$$
and the equation (4-12) is reduced to 
the Konopelchenko's Dirac equation (1-1) [3-6],
which represents the symmetry of the immersed surface in $\Bbb R^3$; 
$$
        \pmatrix  p_1 & \partial \\
                 \bar \partial & p_1  \endpmatrix
                   \pmatrix \psi_{a+}\\ \psi_{a-} \endpmatrix=0. 
               \tag 5-6
$$
This is one of the Lax operator of MNV equation [3-7].
Thus the Dirac operator in (4-12) contains my previous result in 
ref.[15] as its special case.

When I use the complex parameterization of the part of the 
euclidean space,
$$
	Z :=x^1 +\ii x^2,  \tag 5-7
$$
we have  special solutions of (5-6) [3-6],
$$
	2 \ii (\psi_{a+})^2:=-\bar \partial \bar Z, \quad
	2 \ii (\psi_{a-})^2:= \partial \bar Z,\quad
	-2  \psi_{a+}\psi_{a-}= \partial x^3 , \tag 5-8
$$ 

Hence the Dirac operator in (4-12) partially exhibits the symmetry 
of immersed surface.

\subheading{\S 5-3 Reduction to a minimal surface in $\Bbb R^3$}

Provided that $H_\xi=0$, the equation of motion (4-12)  
becomes the original Weierstrass-Enneper formula [1-6,15]
which was obtained as a scheme to obtain a minimal surface
in last century and the lagrangian (4-10) becomes that
in the ordinary (classical) conformal field theory [17].

Thus the Dirac operator in (4-12) contains
the original Weierstrass-Enneper formula.

\tvskip
\centerline{\twobf \S 6  Discussion} 
\tvskip

In this article, I obtained the Dirac operator defined over a 
conformal surface $\SS$ immersed in $\Bbb R^4$. 
As I argued in \S 5, the Dirac operator in (4-12) partially 
exhibits the symmetries of the immersed object;
it recovers the Lax operators of the integrable 
surface and curve by taking appropriate limits.
Thus I believe that it is natural and can be partially
regarded as a generalization of the Dirac 
operator of Konopelchenko or that of the generalized Weierstrass 
equation. Consequently I conjecture that it would exhibit 
the symmetry of such immersion in $\Bbb R^4$.
This conjecture could not be proved in this stage but if true,
using this operator, one can quantize a surface in $\Bbb R^4$
as I did in ref.[21].

Finally I will comment upon the higher dimensional case.
As I did in ref.[10], it  is not so difficult that
 the argument in this article can be generalized to a case of a 
 surface in more higher
dimensional space $\Bbb R^n$ $n>3$.
Thus it is expected that further studies on this Dirac operator 
might be important for the immersion and quantization 
of a surface in $\Bbb R^{10}$ or $\Bbb R^{24}$.

\tvskip
\centerline{\twobf Acknowledgment}
\tvskip

I would like to thank  Y.~\^Onishi and Prof.~S.~Saito for 
critical discussions and  continuous encouragements
and Prof.~B.~G.~Konopelchenko and Prof.~I.~A. 
Taimanov for helpful comment,
sending me their interesting works and encouragement.
I am grateful Prof.~P.~G.~Grinevich  for sending me his 
interesting works.
I also acknowledge that the seminars on
differential geometry, topology, knot theory and group theory with
Prof.~K.~Tamano influenced this work.



\Refs
\ref \no 1 \by A.~I.~Bobenko \jour Math.~Ann. \vol 290 \pages 209-245
\yr 1991 \endref

\ref \no 2 \by A.~I.~Bobenko \paper
Surfaces in terms of 2 by 2 matrices: Old and new integrable cases
\inbook Harmonic Maps and Integrable Systems 
\eds A.~P.~Fordy and J.~C.~Wood
\publ Vieweg \publaddr Wolfgang Nieger \yr 1994 \endref

\ref \no 3 \by B.~G.~Konopelchenko  \jour Studies in Appl.~Math.   
\vol 96  \yr1996 \page9-51 \endref

\ref \no 4 \by B.~G.~Konopelchenko and I.~A.~Taimanov
 \jour J.~Phys.~A: Math.~\& Gen.  
\vol 29  \yr1996 \page1261-65 \endref



\ref \no 5 \by  I.~A.~Taimanov  
\paper Modified Novikov-Veselov equation and 
differential geometry of surface \jour dg-ga/9511005  \endref


\ref \no 6 \by  I.~A.~Taimanov  
\paper The Weierstrass representation of closed surfaces 
in $\Bbb R^3$ \jour dg-ga/9710020  \endref

\ref \no 7 \by   A.~P.~Veselov and S.~P.~Novikov 
 \jour Soviet Math.~Dokl. \vol 30 \pages 588-591, 705-708 
 \yr 1984   \endref

\ref \no 8 \by S. Matsutani and H.~Tsuru \jour Phys. Rev A 
\vol 46 \yr1992\page1144-7 \endref

\ref \no 9 \by S.~Matsutani  \jour Prog. Theor. Phys. \vol 91 
\yr1994\page1005-37 \endref

\ref \no 10 \by S.~Matsutani  \jour Phys. Lett. A.   \vol 189 
\yr1994 \page27-31 \endref

\ref \no 11 \by S.~Matsutani \jour  J.  Phys. A: Math. \& Gen.
 \vol 28 \yr 1995 \page 1399-1412 \endref

\ref \no 12 \by S.~Matsutani \jour Int. J. Mod. Phys. A \vol 10 
\yr 1995 \page 3091-3107 \endref

\ref \no 13 \by S.~Matsutani \book Thesis in Tokyo Metropolitan Univ. 
\yr 1995  \endref

\ref \no 14 \by M.~Burgess and B.~Jensen   \jour Phys. Rev. A    
\vol48  \yr1993\page1861-6 \endref

\ref \no 15 \by S.~Matsutani \jour  J.  Phys. A: Math. \& Gen. 
\vol 30 \yr 1997 \page 4019-4029 \endref

\ref \no 16 \by S.~Matsutani 
\paper Immersion Anomaly of Dirac Operator on Surface in $\Bbb R^3$ 
\jour physics9707010 \endref

\ref \no 17 \by A.~M.~Polyakov  \book Gauge Fields and Strings
\publ Harwood Academic Publishers \yr 1987 \publaddr London \endref


\ref \no 18 \by R.~Carroll and B.~Konopelchenko 
\jour Int. J. Mod. Phys. \vol A11 \yr 1996 \pages1183-1216
\endref

\ref \no 19 \by P.~G.~Grinevich and M.~U.~Schmidt \paper 
Conformal invariant functionals of immersioons of tori 
into $\Bbb R^3$ \jour dg-ga/9702015 \endref
\ref \no 20 \by B.~G.~Konopelchenko and G.~Landolfi 
\paper On classical string configurations 
 \jour solv-int9710008 \endref
\ref \no 21 \by S.~Matsutani  \paper 
Quantization of Willmore surface $\Bbb R^3$ 
\jour solv-int/9707007 \endref

\ref \no 22  \by H.~Hashimoto
\yr 1972 \jour J. Fluid Mech. \vol 51\pages 477-85\endref

\ref \no 23  \by R.~E.~Goldstein and D.~M.~Petrich
\jour Phys. Rev. Lett.\vol  67  \yr 1991 \page 3203-3206 \endref
\ref \no 24\by J.~Langer and R.~Perline 
 \jour  J.~Nonlinear Sci.  \vol 1 \yr 1991 \page 71-91 \endref
\ref \no 25\by A.~Doliwa and P.~M.~Santini \jour  Phys.~Lett.~A  \vol 
185 \yr 1994 \page 373-384 \endref
\ref \no 26 \by S.~Matsutani \paper 
 \jour to appear J.~Phys.~A / solv-int/9707003 \endref
\ref \no 27 \by S.~Matsutani \paper 
 \jour in preparation \endref

\ref \no 28 \by P.~G.~Grinevich and M.~U.~Schmidt
\paper Closed curves in $\Bbb R^3$:
 a characterization in terms of curvature and torsion, the Hasimoto
 map and periodic solutions of the Filament Equation
\jour dg-ga9703020 \endref
 
\ref \no 29 \by S.~Kobayashi  \book Differential Geometry
of  Curves and Surfaces, second ed. (in Japanese)
\publ Shokabo \yr 1995 \publaddr Tokyo \endref

\ref \no 30 \by R.~C.~T.~da Costa  \jour Phys. Rev. A    \vol23  
\yr1981 \page1982-7 \endref

\ref \no 31  \by H.~Jensen  and H.~Koppe   \jour Ann. Phys.    
\vol 63 \yr1971 \page586-91 \endref

\ref \no 32  \by P.~Duclos, P.~Exner and 
P.~$\overset\smallsmile\to{\text{S}}
\overset\smallsmile
\to{\text{t}}$ov{\'i}$\overset\smallsmile\to{\text{c}}$ek
\jour Ann.~Inst.~Henri Poincar\'e \vol 62 \yr 1995 
\pages 81-101 \endref

\ref \no 33 \by  P.~B.~Gilkey \book Invariance Theory, 
The Heat Equation and
the Atiyah-Singer Index Theorem
\publ Publish or Perish \yr 1984 \publaddr Wilmington \endref
 
\ref \no 34 \by  M.~J.~Ablowitz, D.~J.~Kaup, A.~C.~Newell
 and H.~Segur
\jour Stud.~Appl,~Math. \vol 53 \pages 249-315 \endref



\endRefs
\enddocument